\documentclass[journal=jpclcd,manuscript=article]{achemso}
\setkeys{acs}{articletitle=true}

\usepackage[version=3]{mhchem} 

\usepackage{amsmath}
\usepackage{amsfonts}
\usepackage{amssymb}
\usepackage{graphicx}
\usepackage{dcolumn}
\usepackage{bm}
\usepackage{subcaption}
\usepackage{algorithm} 
\usepackage{algpseudocode} 
\usepackage{xr}
\usepackage{hyperref}
\usepackage{braket}
\usepackage{xcolor}

\makeatletter
\newcommand*{\addFileDependency}[1]{
  \typeout{(#1)}
  \@addtofilelist{#1}
  \IfFileExists{#1}{}{\typeout{No file #1.}}
}
\makeatother

\usepackage[capitalize]{cleveref}
\crefname{figure}{Fig.}{Figs.}
\Crefname{figure}{Figure}{Figures}
\crefname{table}{Tab.}{Tabs.}
\Crefname{table}{Table}{Tables}
\crefname{equation}{Eq.}{Eqs.}
\Crefname{equation}{Equation}{Equations}
\crefname{section}{Sec.}{Secs.}
\Crefname{section}{Section}{Sections}
\usepackage{xcolor}

\usepackage{array}
\newcommand{\PreserveBackslash}[1]{\let\temp=\\#1\let\\=\temp}
\newcolumntype{C}[1]{>{\PreserveBackslash\centering}p{#1}}
\newcolumntype{R}[1]{>{\PreserveBackslash\raggedleft}p{#1}}
\newcolumntype{L}[1]{>{\PreserveBackslash\raggedright}p{#1}}

\usepackage{multirow}

\SectionNumbersOn

\author{Alessio Fallani}
\email{alessio.fallani@algorithmiq.fi}
\affiliation{\fontsize{10pt}{10pt}\selectfont Algorithmiq Ltd., Kanavakatu 3C, FI-00160 Helsinki, Finland}

\author{Pi A. B. Haase}
\affiliation{\fontsize{10pt}{10pt}\selectfont Algorithmiq Ltd., Kanavakatu 3C, FI-00160 Helsinki, Finland}

\author{Julianne F. F. Eckert}
\affiliation{\fontsize{10pt}{10pt}\selectfont Algorithmiq Ltd., Kanavakatu 3C, FI-00160 Helsinki, Finland}

\author{Luukas Nikkanen}
\affiliation{\fontsize{10pt}{10pt}\selectfont Algorithmiq Ltd., Kanavakatu 3C, FI-00160 Helsinki, Finland}
\altaffiliation{\fontsize{10pt}{10pt}\selectfont Department of Chemistry, University of Helsinki, P.O. Box 55, FI-00014 University of Helsinki, Finland}

\author{Sherri A. McFarland}
\affiliation{\fontsize{10pt}{10pt}\selectfont Department of Chemistry and Biochemistry, The University of Texas at Arlington, Arlington, Texas 76019, United States}

\author{Martina Stella}
\affiliation{\fontsize{10pt}{10pt}\selectfont Algorithmiq Ltd., Kanavakatu 3C, FI-00160 Helsinki, Finland}
\altaffiliation{\fontsize{10pt}{10pt}\selectfont The Abdus Salam International Centre for Theoretical Physics (ICTP), Condensed Matter and Statistical Physics Section, Trieste 34151, Italy}

\author{Fabijan Pavo\v{s}evi\'{c}}
\email{fabijan.pavosevic@algorithmiq.fi}
\affiliation{\fontsize{10pt}{10pt}\selectfont Algorithmiq Ltd., Kanavakatu 3C, FI-00160 Helsinki, Finland}

\title[]
  {Data-Efficient Active Learning Discovery of Transition Metal Photosensitizers for Type I Photodynamic Therapy}

\begin{document}



\begin{tocentry}
\begin{figure}[H]
	\begin{center}
		\includegraphics[width=2.0in]{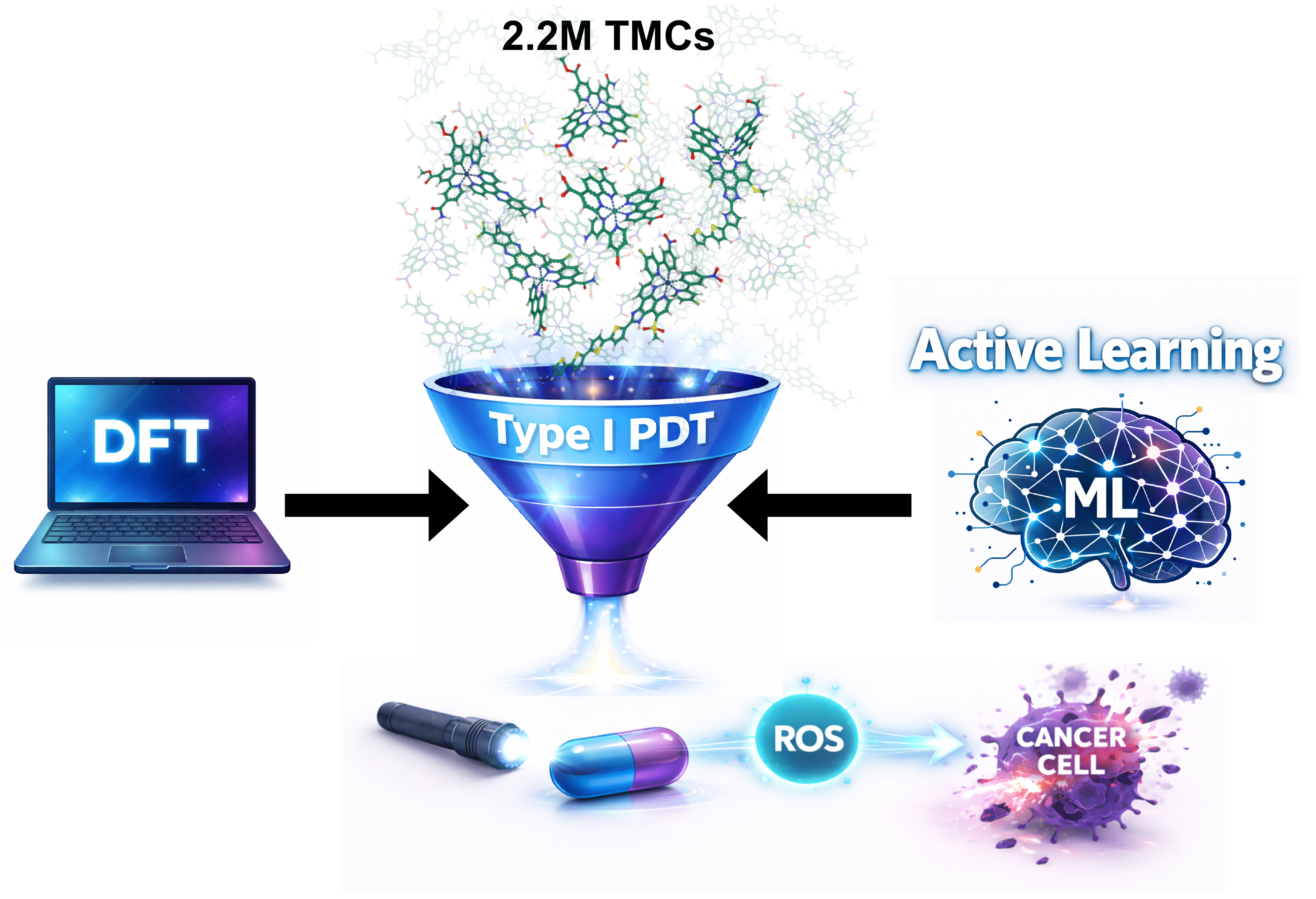}
	\end{center}
\end{figure}
\end{tocentry}

\begin{abstract}

Transition-metal complexes (TMCs) are promising photosensitizers for Type~I photodynamic therapy (PDT), where electron-transfer processes can generate reactive oxygen species under hypoxic conditions. Yet identifying candidates with the required ground- and excited-state redox energetics remains challenging across the vast chemical space of TMCs. Here, we develop a data-efficient active learning (AL) framework for the discovery of Type~I active TMC photosensitizers by combining a chemically structured design space of over 2.1 million Ru(II), Os(II), and Ir(III) complexes with targeted DFT calculations and pretrained atomistic representations. With only 300 quantum-chemical evaluations, the approach efficiently enriches candidates within a mechanistically defined optimal redox region. Analysis of the viable complexes reveals chemical design principles linking metal identity, ligand framework, substituent pattern, and physicochemical properties to Type~I photoreactivity, including a pronounced preference for Os(II)-based complexes and electronically asymmetric ligand environments along with combination of electronic donating and accepting substituents. More broadly, the strategy presented herein provides a scalable, mechanism-guided route for the rational design of transition-metal photocatalysts for applications spanning biomedicine, solar energy conversion, and photoredox chemistry.

\end{abstract}

\maketitle

\section{Introduction}

\begin{figure*}[ht!]
  \centering
  \includegraphics[width=6.5in]{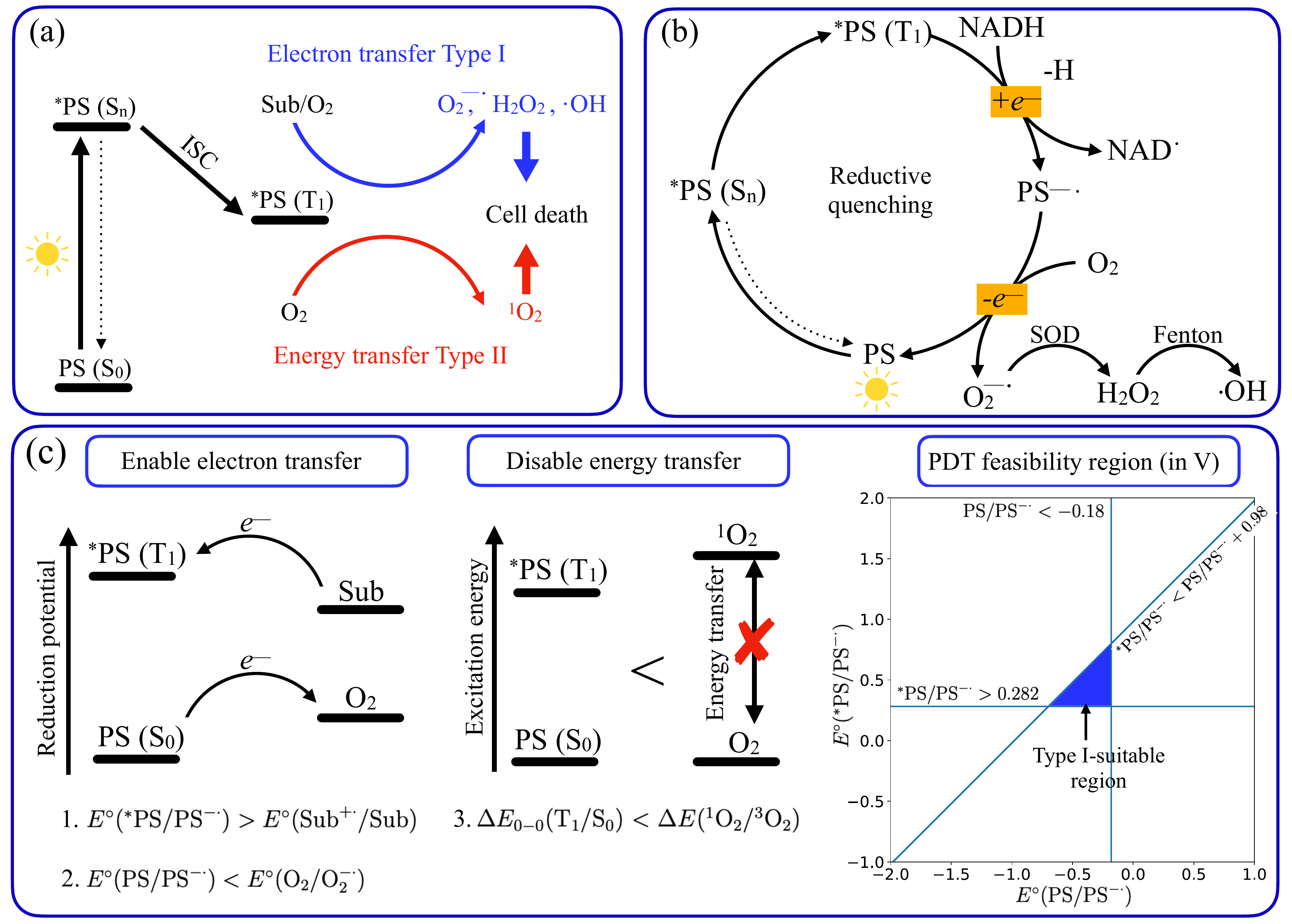}
  \caption{(a) Schematic illustration of the Type~I and Type~II PDT mechanisms. (b) Detailed schematic illustration of catalytic cycle of the Type~I PDT working mechanism. (c) Reduction potential requirements and triplet energy requirements for the design of Type~I PDT compounds. Objective function for search of optimal (blue triangle area) Type~I PDT agents.}
  \label{fig:Figure1}
\end{figure*}

Photodynamic therapy (PDT) is a clinically established treatment that has emerged as a valuable alternative or complement to surgery, chemotherapy, and radiotherapy, particularly for nonresectable or localized tumors.~\cite{dolmans2003photodynamic,agostinis2011photodynamic} Owing to its minimal invasiveness, reduced systemic toxicity, low propensity to induce drug resistance, and exceptional spatiotemporal selectivity, PDT has expanded from early dermatological applications to a broad range of oncological indications and antimicrobial therapies.~\cite{dolmans2003photodynamic,agostinis2011photodynamic,ran2023photocatalytic,obaid2024engineering,cai2025phototherapy} The therapeutic effect of PDT arises from light activation of a photosensitizer (PS), which generates cytotoxic reactive oxygen species (ROS) upon interaction with molecular oxygen. The efficiency and identity of the ROS produced are governed by the excited-state properties of the PS. Upon light absorption, the PS is promoted from its ground state ($\text{S}_0$) to excited singlet states ($\text{S}_{\text{n}}$), which can undergo intersystem crossing (ISC) to populate a long-lived triplet state ($\text{T}_1$) responsible for photochemical reactivity (see Fig~\ref{fig:Figure1}a).~\cite{turro2009principles,schweitzer2003physical,baptista2017type} From this triplet state ($^*\text{PS}$), ROS generation proceeds via two principal mechanisms: Type~II and Type~I.~\cite{foote1991definition} In the Type~II pathway, the $^*\text{PS}$ transfers energy to molecular oxygen to produce singlet oxygen, which is the dominant mechanism for most clinically used photosensitizers, particularly porphyrin and chlorin derivatives.~\cite{dolmans2003photodynamic,cai2025phototherapy} However, the efficiency of this pathway is intrinsically limited by oxygen availability. Many solid tumors are poorly vascularized and contain extensive hypoxic regions, leading to reduced singlet oxygen generation and diminished therapeutic effectiveness.~\cite{brown2004exploiting,wen2022achieving}

By contrast, Type I photochemical pathways (Fig~\ref{fig:Figure1}a and Fig~\ref{fig:Figure1}b) involve electron- or hydrogen-transfer reactions originating from $^*\text{PS}$.~\cite{baptista2017type} The $^*\text{PS}$ can interact with nearby biomolecular substrates (Sub) such as nicotinamide adenine dinucleotide (NADH), glutathione (GSH), redox-active amino acid residues, unsaturated membrane lipids, or nucleic acid bases, generating radical intermediates on either the PS or the substrate.~\cite{teng2025rational,li2025new} These radicals can subsequently react with molecular oxygen to produce ROS such as superoxide, or participate in further radical reactions that damage nearby biomolecules. In some cases, superoxide may undergo superoxide dismutase-mediated (SOD) disproportionation to form hydrogen peroxide,~\cite{li2018near,li2022type} which can participate in further redox reactions that generate additional oxidizing species depending on the local biochemical environment.~\cite{li2025new} Because these pathways proceed through electron-transfer chemistry involving local substrates and radical intermediates, Type I photochemical processes may remain operative under conditions where singlet oxygen sensitization is less efficient. This possibility has motivated increasing interest in developing PSs that favor electron-transfer-mediated photochemical pathways.~\cite{li2022type,wang2024recent,teng2025rational,ma2025current,li2025new,vigueras2025insights} However, designing PSs that selectively favor Type I reactivity requires satisfying a narrow and often competing set of excited-state and redox constraints, making rational discovery highly challenging.

Transition-metal complexes provide versatile platforms for the design of next-generation PSs because their photophysical and redox properties can be systematically tuned through ligand engineering. A prominent example is the Ru(II) polypyridyl complex TLD-1433,~\cite{monro2018transition,mcfarland2020metal} which has advanced to Phase II clinical trials for the treatment of non-muscle-invasive bladder cancer.~\cite{lilge2020evaluation,lilge2020minimal,kulkarni2022phase} TLD-1433 is administered as a racemic Ru(II) complex containing two 4,4$'$-dimethyl-2,2$'$-bipyridyl ligands and an imidazo[4,5-\textit{f}][1,10]phenanthroline ligand functionalized with an $\alpha$-terthiophene chromophore. Upon green-light excitation, the complex efficiently populates long-lived triplet excited states capable of sensitizing singlet oxygen with high quantum yield, but it can also participate in electron-transfer reactions.
The oligothienyl substituent is key and introduces intraligand charge-transfer character that enables additional redox photochemistry beyond what is typically observed for simpler Ru(II) polypyridyl complexes. In particular, when the energy of the triplet intraligand charge-transfer ($^3$ILCT) state lies below that of the lowest metal-to-ligand charge-transfer ($^3$MLCT) state, the excited-state manifold supports both energy-transfer and electron-transfer pathways. As a result, TLD-1433 and related oligothienyl-appended Ru(II) complexes~\cite{sun2024hypoxia,cole2024ru,cole2023ru,roque2022intraligand,cole2021anticancer,roque2020ii,roque2020breaking} can generate ROS through multiple mechanisms and have been shown to retain significant phototoxic activity under hypoxic conditions. These findings illustrate how careful control of excited-state energetics in transition-metal complexes can enable photophysical processes beyond conventional singlet-oxygen sensitization. 

At the same time, the chemical design space associated with transition-metal photosensitizers is extremely large. Small changes in ligand structure can significantly alter absorption properties, excited-state energetics, charge-transfer character, and ground- and excited-state redox potentials. As illustrated by TLD-1433 and related oligothienyl-modified Ru(II) complexes,~\cite{monro2018transition,cole2024ru,cole2023ru,roque2022intraligand,cole2021anticancer,roque2020ii,roque2020breaking} these parameters strongly influence whether a photosensitizer favors singlet-oxygen sensitization, electron-transfer photochemistry, or a combination of both. However, exploring this multidimensional design space through traditional experimental screening alone is inherently slow and resource intensive. These challenges motivate the development of computational strategies capable of rapidly identifying promising photosensitizer candidates with targeted excited-state and redox properties, with quantum chemical methods playing a central role in this effort.

Consequently, quantum chemical methods—particularly density functional theory (DFT) and time-dependent DFT (TD-DFT)—have become indispensable for understanding and predicting the electronic structure and photophysical behavior of transition metal complexes.~\cite{demissie2015dft,latouche2015td,belletto2024detailed} These approaches provide access to key properties such as excitation energies, absorption spectra, singlet–triplet gaps, charge-transfer character, and ground- and excited-state redox potentials,~\cite{escudero2013unveiling,demissie2015dft,latouche2015td,alberto2016theoretical,atkins2017trajectory,alberto2020theoretical,ponte2023computational,spiegel2023tuning,belletto2024detailed,barretta2024computational,barretta2024rational,fennes2024rational,hagmeyer2025dft,zehr2025quantum} and have been widely used in PDT to rationalize Type~I/Type~II selectivity, predict ROS-generating capability, evaluate deactivation pathways, and guide ligand design in Ru(II) and Ir(III) complexes.~\cite{alberto2016theoretical,alberto2020theoretical,spiegel2023tuning,barretta2024rational,barretta2024computational,ponte2023computational} However, the combinatorial chemical space accessible through ligand substitution is enormous and cannot be exhaustively explored at the required level of theory. This challenge, commonly encountered in drug discovery and materials science as high-throughput virtual screening (HTVS),~\cite{curtarolo2013high,vogiatzis2018computational,HTVSrev2,Gomez-Bombarelli2016} requires strategies that efficiently identify promising candidates within vast chemical spaces under limited computational budgets.

Machine learning (ML) approaches are well suited to this problem as they can learn structure--property relationships, enabling rapid estimation of molecular properties across large chemical spaces at a fraction of the cost of quantum chemical calculations.~\cite{bigi2024wigner,Batzner2022,MLelectrocatalysis,cryst15110925,Deng2023molpred,HTVSrev,NEMATOV2025e01139,screeningreviewOliveira,Dreiman2021,Dral25Omnip2x,macefoundation,GEMnetBio,alignnmaterials,Microsoft} These methods can be integrated into discovery workflows in different ways, including as predictive filters to reduce candidate space,~\cite{Microsoft} as generative models to design molecules with targeted properties,~\cite{ReinventAL} or through optimization strategies based on Bayesian optimization (BO), which provides a framework for sequential selection of promising candidates under uncertainty.~\cite{MolPal,Hase2018} This last class, often implemented as active learning (AL),~\cite{bradford2018efficient,smith2018less} iteratively refines surrogate models by selecting the most informative and promising candidates for high-fidelity evaluation, thereby balancing exploration and exploitation.~\cite{bradford2018efficient,smith2018less} By focusing computational effort on these targeted regions of chemical space, AL significantly improves screening efficiency and is particularly well suited for large transition metal complex (TMC) libraries, where each quantum chemical calculation is costly. Recent studies have demonstrated that AL can efficiently navigate million-scale TMC spaces while maintaining predictive reliability.~\cite{janet2020accurate,duan2022active,nandy2021computational,nandy2022new}

In this work, we develop an active learning–driven machine learning framework for the discovery of transition-metal photosensitizers tailored to Type~I PDT. We construct and explore a design space of over 2 million Ru(II), Os(II), and Ir(III) complexes with systematically varied ligand environments, targeting candidates that satisfy mechanistically defined constraints in excited-state and ground-state redox potentials while suppressing competing Type~II pathways. By iteratively coupling ML predictions with targeted DFT calculations, we efficiently screen this space using only 300 quantum chemical evaluations, demonstrating navigation of an otherwise intractable chemical landscape. This data-efficient strategy leverages pretrained atomistic representations, specifically the UMA (Universal Models for Atoms) foundation model,~\cite{kim2025leveragingneuralnetworkinteratomic,ceriottilatentfoundation,cao2023large,Masood2025,alignnmaterials,jang2026boltzstrongbaselineatomlevel,UMA} enabling rapid discovery of complexes intrinsically biased toward Type~I reactivity. More broadly, this approach transforms quantum-chemical screening into a scalable, mechanism-guided discovery platform with applications extending to the discovery of organic and transition-metal photocatalysts for solar fuel production,~\cite{tran2012recent} CO$_2$ reduction,~\cite{ulmer2019fundamentals} water splitting,~\cite{nishioka2023photocatalytic} and photoredox organic synthesis.~\cite{reischauer2021emerging,prier2013visible}

\section{Results and Discussion}
\subsection{Mechanistic Design Criteria for Type I PDT}
The discovery of Type~I PDT agents can be formulated as a constrained multi-criteria optimization problem in which thermodynamic and kinetic requirements for electron-transfer reactivity must be satisfied simultaneously. In  Type~I, the excited photosensitizer ($^*\text{PS}$) engages in electron–transfer reactions with surrounding biomolecular substrates, generating a reduced photosensitizer species ($\text{PS}^{-\cdot}$) that subsequently transfers an electron to triplet oxygen to produce superoxide ($\text{O}_2^{-\cdot}$). The design objective is therefore to identify compounds that preferentially operate through this electron-transfer pathway while suppressing competing Type~II energy-transfer mechanisms. Such design rationale is outlined in Ref.~\citenum{teng2025rational}. In this work, we explicitly implement these criteria and recast them into a computational objective function suitable for high-throughput screening. All of the design criteria are schematically illustrated in Fig~\ref{fig:Figure1}c.

The first criterion is thermodynamic feasibility of biomolecular substrate oxidation by $^*\text{PS}$ as
\begin{equation}
    \label{eqn:PS_Sub}
    {}^{*}\mathrm{PS} + \mathrm{Sub} \rightarrow \mathrm{PS}^{-\cdot} + \mathrm{Sub}^{+\cdot}
\end{equation}
For reductive Type~I pathways, the excited-state reduction potential must be sufficiently positive to abstract an electron from biomolecular substrates, i.e., $E^\circ(^*\text{PS}/\text{PS}^{-\cdot})>E^\circ(\text{Sub}^{+\cdot}/\text{Sub})$. In this work, the substrate redox couple ($\mathrm{Sub}^{+\cdot}/\mathrm{Sub}$) is represented by the $\mathrm{NAD}^{\cdot}/\mathrm{NADH}$ couple, as it has been identified as an important biological redox mediator in several recent studies,~\cite{huang2019targeted,liu2023concerted,wang2024recent,li2025new} although the present computational framework can be readily extended to any other relevant biomolecular redox couple. The $\mathrm{NAD}^{\cdot}/\mathrm{NADH}$ redox couple experimental reduction potential is 0.282~V relative to the standard hydrogen electrode (SHE);~\cite{anderson1980energetics} therefore, the suitable photosensitizer in excited state should satisfy $E^\circ(\mathrm{^*PS}/\mathrm{PS}^{-\cdot}) > 0.282~\mathrm{V}$ vs. $\mathrm{SHE}$.

The second requirement is compatibility with oxygen reduction. Following substrate quenching, the reduced photosensitizer $\text{PS}^{-\cdot}$ must be capable of transferring an electron to triplet oxygen to form superoxide as
\begin{equation}
    \label{eqn:PS_O2}
    \mathrm{PS}^{-\cdot} + \mathrm{O}_2 \rightarrow \mathrm{PS} + \mathrm{O}_2^{-\cdot}
\end{equation}
Accordingly, the ground-state reduction potential of the photosensitizer must be appropriately positioned relative to the $\text{O}_2/\text{O}_2^{-\cdot}$, such that $E^\circ(\text{PS}/\text{PS}^{-\cdot})<E^\circ(\text{O}_2/\text{O}_2^{-\cdot})$. The standard reduction potential of the $\mathrm{O_2}/\mathrm{O_2}^{-\cdot}$ couple has been experimentally determined to be $-0.18~\mathrm{V}$ vs. $\mathrm{SHE}$;~\cite{koppenol2010electrode} therefore, a suitable photosensitizer must satisfy $E^\circ(\mathrm{PS}/\mathrm{PS}^{-\cdot}) < -0.18~\mathrm{V}$ vs. $\mathrm{SHE}$ to enable thermodynamically favorable electron transfer to oxygen.

The third design condition is suppression of Type~II energy transfer. Efficient singlet oxygen production requires the triplet energy of the photosensitizer to exceed the excitation energy of molecular oxygen. Therefore, Type~I selectivity can be promoted by designing photosensitizers with triplet energies below the $^1\text{O}_2$ sensitization threshold, rendering energy transfer thermodynamically unfavorable.~\cite{ma2025current} This condition is achieved if $\Delta E_{0-0}(\text{T}_1/\text{S}_0) < \Delta E(^1\text{O}_2/^3\text{O}_2)$, where the $\Delta E_{0-0}(\text{T}_1/\text{S}_0)$ is energy of adiabatic $\text{T}_1 \rightarrow \text{S}_0$
transition and $\Delta E(^1\text{O}_2/^3\text{O}_2)$ is energy of singlet oxygen sensitation. Because the energy required for the formation of singlet oxygen in aqueous solution is  $0.98~\mathrm{eV}$,~\cite{herzberg1950spectra} a photosensitizer designed to operate predominantly through a Type~I mechanism should possess $\Delta E_{0-0}(\text{T}_1/\text{S}_0) < 0.98~\mathrm{eV}$ in order to disfavor Type~II energy transfer.

These three constraints on 
$E^\circ(\mathrm{PS}/\mathrm{PS}^{-\cdot})$, 
$E^\circ({}^*\mathrm{PS}/\mathrm{PS}^{-\cdot})$, and $\Delta E_{0-0}(\mathrm{T}_1/\mathrm{S}_0)$ along with the Rehm-Weller expression,~\cite{rehm1970kinetics} $E^\circ({}^*\mathrm{PS}/\mathrm{PS}^{-\cdot})=E^\circ(\mathrm{PS}/\mathrm{PS}^{-\cdot})+\Delta E_{0-0}(\mathrm{T}_1/\mathrm{S}_0)$, can be reduced after algebraic rearrangement to the following pair of inequalities
\begin{equation}
    \label{eqn:Cond1}
    -0.698~\mathrm{V} < E^\circ(\mathrm{PS}/\mathrm{PS}^{-\cdot}) < -0.18~\mathrm{V}
\end{equation}
and
\begin{equation}
    \label{eqn:Cond2}
    0.282~\mathrm{V} < E^\circ({}^*\mathrm{PS}/\mathrm{PS}^{-\cdot}) 
< 0.98~\mathrm{V} + E^\circ(\mathrm{PS}/\mathrm{PS}^{-\cdot})
\end{equation}
These relations define a triangular region in the 
$\big(E^\circ(\mathrm{PS}/\mathrm{PS}^{-\cdot}), 
E^\circ({}^*\mathrm{PS}/\mathrm{PS}^{-\cdot})\big)$-plane as shown in the right end of Fig~\ref{fig:Figure1}c. 
A photosensitizer with redox potentials located within this region is thus expected to exhibit optimal performance for Type~I PDT.

To evaluate a physicochemical parameter relevant to subcellular localization and PDT efficacy, we computed the logarithm of the partition coefficient (log$P$) as a descriptor of lipophilicity. Lipophilicity has been shown to correlate with membrane permeability, intracellular accumulation of Ru(II) photosensitizers, and with the cytotoxicity.~\cite{mitchell2025photodynamic} In our screening framework, log$P$ was also included as a descriptor that balances aqueous solubility with sufficient hydrophobic character for cellular uptake.~\cite{roque2020strained,mitchell2025photodynamic}
Computationally, log$P$ was obtained using a thermodynamic cycle based on solvation Gibbs free energies in water and 
n-octanol. The standard free energy of transfer from water (w) to octanol (o) was evaluated as
\begin{equation}
\Delta G^\circ_{\text{o/w}} = G^\circ_{\text{oct}} - G^\circ_{\text{water}}
\end{equation}
where $G^\circ_{\text{oct}}$ and $G^\circ_{\text{water}}$ correspond to the solvation free energies calculated in the respective media. The partition coefficient was then determined from
\begin{equation}
\log P = -\frac{\Delta G^\circ_{\text{o/w}}}{2.303RT}
\end{equation}
with $R$ denoting the gas constant and $T$ the absolute temperature.

\subsection{Large-Scale Design Space of TMC Photosensitizers}
The performance of AL in chemical discovery strongly depends on the composition of the initial training set, as this dataset defines the model’s chemical space coverage and uncertainty estimates. For Type~I PDT, it is therefore crucial to construct a chemically diverse database that captures the key structural, photophysical, and physicochemical  features governing reactive oxygen species generation and cytotoxicity.~\cite{wang2024recent,ma2025current,vigueras2025insights} In these systems, candidate activity is strongly influenced by three primary components: the metal center, the ligand framework, and the ligand substituents. Ensuring sufficient variation across these components enables robust model generalization and more reliable identification of promising candidates.

\begin{figure*}[ht!]
  \centering
  \includegraphics[width=5.0in]{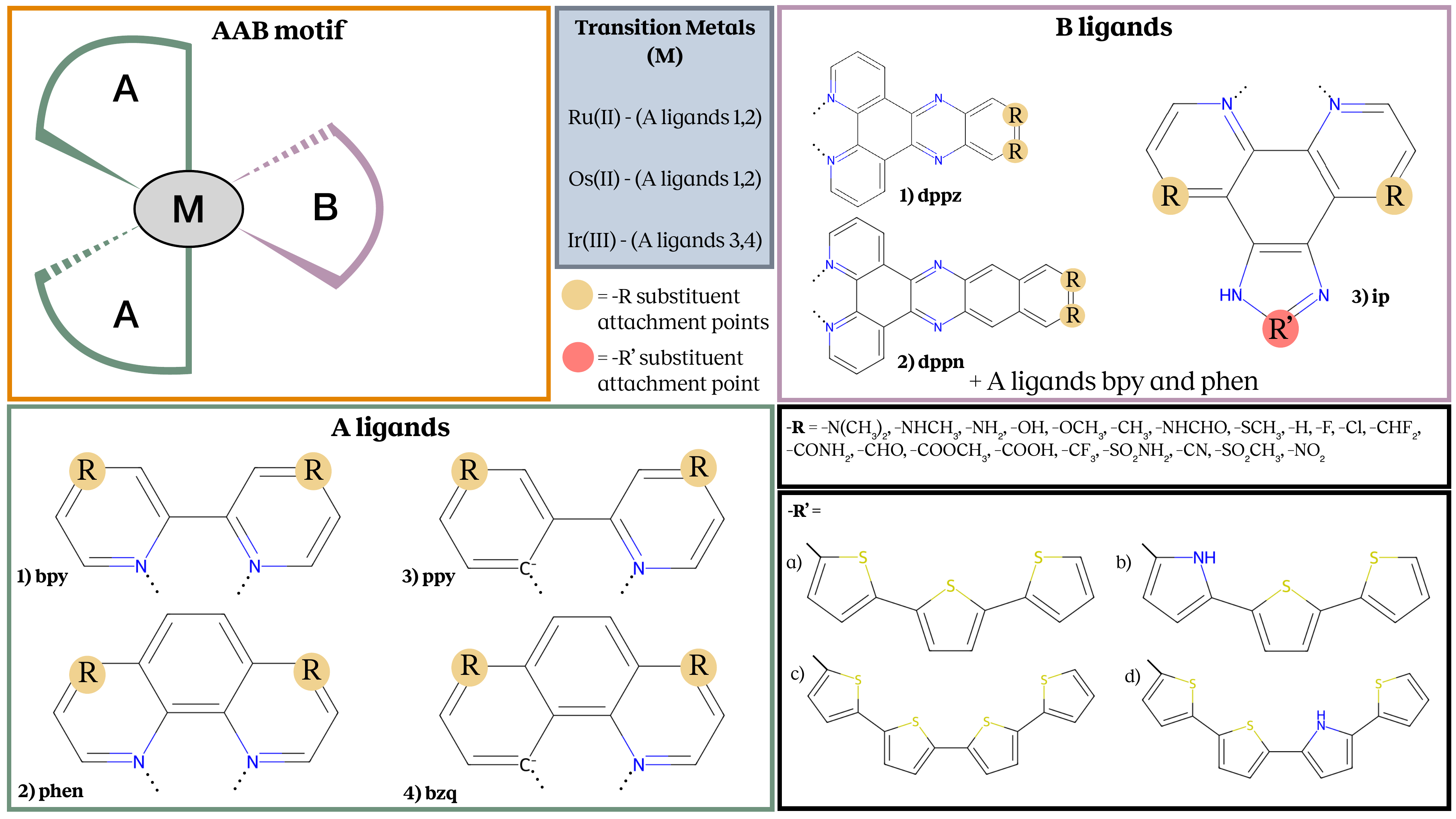}
  \caption{Schematic representation of the database construction strategy. The octahedral AAB scaffold is shown in the orange box, with A ligands (green box), B ligands (purple box), --R and --R' substituents (black boxes), attachment points (yellow and red circles), and investigated metal centers (grey box). --R substituents are ordered from electron donating to withdrawing groups according to $\sigma_\text{p}$ from Ref.~\citenum{hansch1991survey}.}
  \label{fig:DB_strategy}
\end{figure*}

Guided by the scaffold of TLD-1433,~\cite{monro2018transition} we built a database around a general AAB motif in which a transition metal lies at the core of the structure, coordinated by two identical ligands (A) and one distinct ligand (B) in an octahedral arrangement. The overall database strategy is illustrated in Fig.~\ref{fig:DB_strategy}. In this schematic, the scaffold of the octahedral motif is highlighted in the orange box, metal centers in the gray box, A ligands in the green box, B ligands in the purple box, --R and --R$'$ substituents in the black boxes, and yellow and red circles indicate the attachment points for --R and --R$'$ substituents, respectively. As central metals, we considered Ru(II), Os(II), and Ir(III). Due to the heavy-atom effect, Os(II) and Ir(III) exhibit stronger spin–orbit coupling than Ru(II), promoting more efficient intersystem crossing and triplet-state formation. The A ligands---1) 2,2$'$-bipyridine (\textit{bpy}), 2) 1,10-phenanthroline (\textit{phen}), 3) 2-phenylpyridine (\textit{ppy}), and 4) benzoquinoline (\textit{bzq})---were selected for their extended $\pi$-conjugation, structural rigidity, photostability, and strong chelating ability. These features reduce the nonradiative decay and provide control over the nature and energetics of charge-transfer excitations. The B ligands consist of $\pi$-extended bipyridine (\textit{dppz} and \textit{dppn}) and imidazo-phenanthroline (\textit{ip}) ligands. The \textit{bpy} and \textit{phen} ligands were also used as B ligands to increase dataset diversity. Ru(II) and Os(II) were paired with \textit{bpy} and \textit{phen} A ligands, whereas Ir(III)-based complexes were constructed using \textit{ppy} and \textit{bzq} ligands. B ligands were utilized for all three metal atoms. Two families of substituents, denoted --R and --R$'$, were also utilized. The --R groups  (placed at the positions indicated by the yellow circles) were selected with increasing electron-withdrawing power to tune HOMO and LUMO energies. The --R$'$ groups---applied exclusively to \textit{ip} B ligand at the position highlighted by the red circle---consist of three- and four thiophene/pyridine-ring tails. To systematically generate molecular candidates, these three relevant components were combined according to a defined set of design guidelines. This combinatorial strategy ultimately resulted in a database comprising $2~170~434$ candidates, providing a well-structured and chemically diverse molecular space with strong potential for identifying optimized Type~I PDT agents.

\subsection{Active Learning for Targeted Discovery}
The objective of this work is to identify, within a dataset of over two million TMCs, those complexes whose properties fall inside a predefined triangular region in the $\big(E^\circ(\mathrm{PS}/\mathrm{PS}^{-\cdot})$, $ 
E^\circ({}^*\mathrm{PS}/\mathrm{PS}^{-\cdot})\big)$-plane, while performing as few quantum chemical calculations as possible. A brute-force evaluation of the entire chemical space at the chosen level of theory is computationally intractable. Following prior studies that combine machine learning with adaptive sampling strategies for accelerated materials and molecular discovery,~\cite{janet2020accurate,nandy2021computational,nandy2022new} we employ an active learning–driven pipeline. As already mentioned in Introduction, this approach iteratively trains surrogate models on a progressively expanding set of computed examples and uses them to guide the selection of new candidates for quantum chemical evaluation, thereby concentrating computational effort on the most informative and promising regions of the search space.

\begin{figure*}[ht!]
  \centering
  \includegraphics[width=\textwidth]{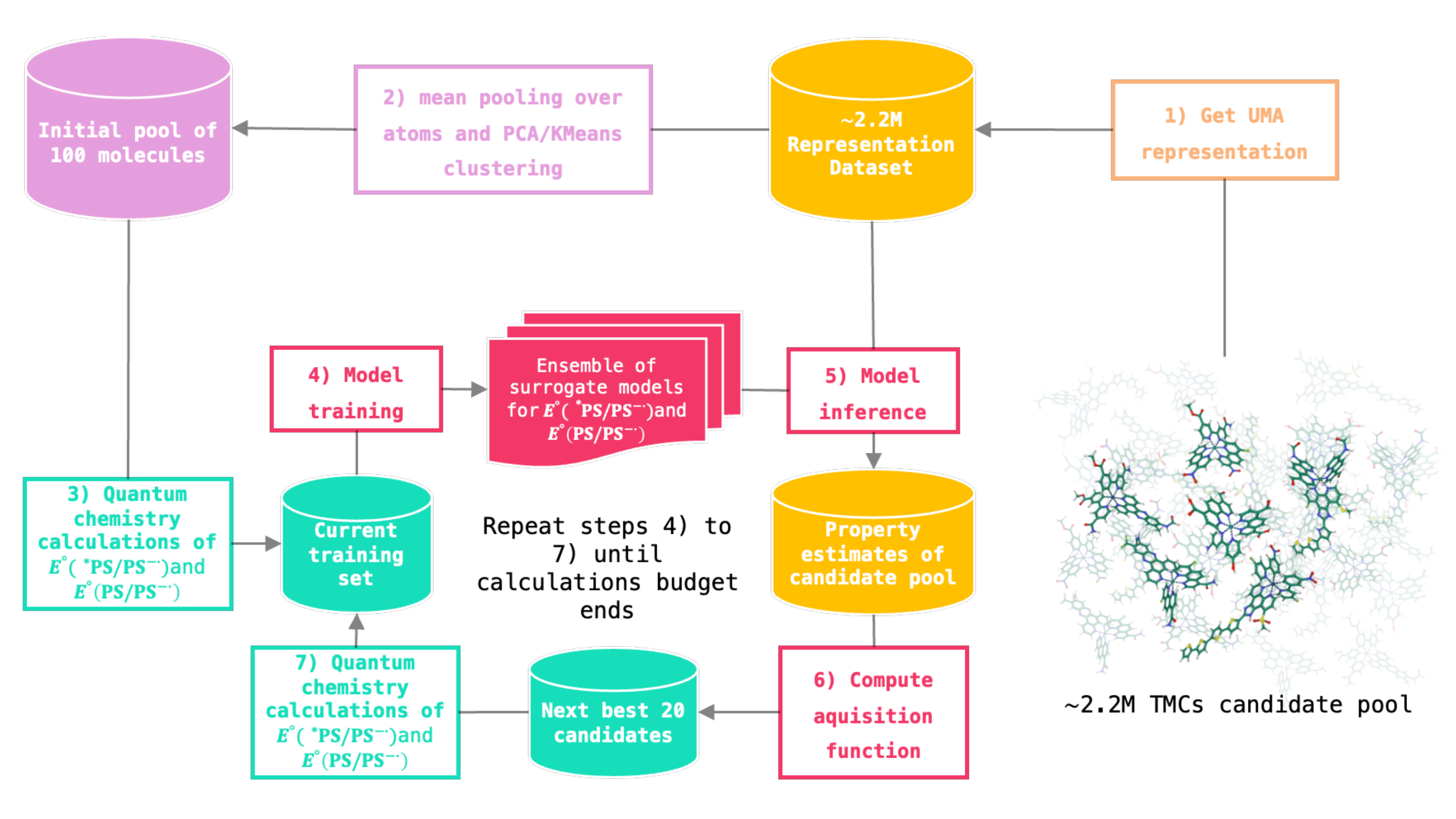}

  \caption{Schematic depiction of the active-learning pipeline. 1) UMA atomic representations are extracted for all structures in the candidate pool and stored in a database. 2) A coarse molecular representation is obtained by mean pooling of atomic representations, followed by PCA dimensionality reduction and KMeans clustering into 100 clusters. The 100 molecules closest to the cluster centroids are selected as the initial training set. 3) Quantum chemistry calculations are performed for these 100 molecules to obtain labels corresponding to $\big(E^\circ(\mathrm{PS}/\mathrm{PS}^{-\cdot}), E^\circ({}^*\mathrm{PS}/\mathrm{PS}^{-\cdot})\big)$. 4) An ensemble of surrogate models is trained on different CV splits of the initial training set. 5) The trained models are used to estimate the properties of the remaining candidate complexes. 6) An acquisition function is computed from these predictions to prioritize the 20 candidates most likely to lie in the optimal region, diversified by cluster. 7) Quantum chemistry calculations are performed for these 20 molecules and the results are added to the training set. Steps 4)--7) are repeated until the computational budget is exhausted.}
  \label{fig:pipeline}
\end{figure*}

Our pipeline, depicted in Fig.~\ref{fig:pipeline}, proceeds in multiple steps. In the first step, all candidate geometries are encoded using the UMA model~\cite{UMA}. To do so, from the last layer we extract an $N_{atoms}\times9\times128$ dimensional representation, retaining only the $N_{atoms}\times1\times128$ invariant component, which corresponds to the scalar ($\ell=0$) channel in the spherical harmonic decomposition, while discarding the higher-order ($\ell>0$) equivariant components. This is consistent with how predictions are made in the original UMA architecture \cite{UMA}, where only the $\ell=0$ component is used in the final layer for prediction. This produces, for each molecule, a 128-dimensional atomic tensor representation trained on DFT energies and forces.~\cite{UMA} The representation is invariant under rotation and encodes global molecular information via multiple layers of message passing. A coarse molecule-level representation is then obtained via average pooling, and dimensionality is reduced from 128 to 20 using PCA, preserving over 99\% of the variance. Batched K-Means clustering\cite{minikmeans} is subsequently applied, with the number of clusters set to be equal to the number of molecules in the initial training set (100); molecules closest to the cluster centroids are selected to form this initial set. Quantum chemical calculations are performed on these 100 structures to obtain the labels $\big(E^\circ(\mathrm{PS}/\mathrm{PS}^{-\cdot}), 
E^\circ({}^*\mathrm{PS}/\mathrm{PS}^{-\cdot})\big)$, forming the initial training set.

Given the initial training set, an ensemble of surrogate models is trained on different cross-validation splits of this set and used to predict the target properties for the remaining molecules. These predictions are used to compute an acquisition function, defined as the product of a utility function inversely proportional to the distance from the triangular region and the Probability of Feasibility ($PoF$), i.e., the fraction of predictions falling inside the target area. This is defined as:
\begin{equation}
\mathcal{S} (x)= \frac{1}{1 + d(x)}\times(1+PoF(x))    
\end{equation}
where $x$ is a place holder for the datapoint considered and $d$ is the expectation of the distance from the triangle over the ensemble surrogate posterior (average from the ensemble predictions) normalized on the maximum estimated value in the current cycle. The addition of 1 to the $PoF$ is aimed at avoiding the flattening of the score for cases where no prediction in the ensemble is in the targeted triangular region. 
This form emphasizes exploitation toward the target region, while tie-breaking among candidates with identical scores uses predictive uncertainty to introduce limited exploration. While conceptually related to standard acquisition functions such as constrained Expected Improvement (EI),~\cite{gardner14} this approach is best understood as an ad-hoc greedy heuristic tailored to the specific model and target setting — rather than a principled acquisition function — as it does not use an improvement term, reflecting that all points within the target region are equally desirable and would saturate such a metric. During each cycle, the best molecule according to this acquisition metric is selected from each of 20 clusters, defined in the same manner as the initial clustering, ensuring some degree of diversity among the sampled molecules. The quantum chemical data of these 20 molecules are then computed and added to the training set. This procedure is repeated until the computational budget is exhausted, which in this case corresponds to 10 cycles, resulting in a total of 300 evaluated molecules (100 initial + 10 × 20 per cycle for a total of 300 calculations).

\subsection{Machine Learning Models}

In active learning–based virtual screening, the choice of molecular representation and surrogate model is critical, as it directly affects compound selection efficiency. Effective models must balance data efficiency, predictive accuracy, reliable uncertainty quantification, and generalization performance. 
In this context classical machine learning methods have been favored for their robustness in low-data regimes \cite{Wojcikowski2017RF}, but deep learning approaches—especially graph neural networks and transformers that leverage on large scale pretraining strategies \cite{Chemeleon,Hu2020Strategies}—have gained traction in HTVS settings due to their greater representational capacity and ability to learn task-specific features end-to-end\cite{MolPal,alignnmaterials}. 

In particular, recent work~\cite{cao2023large} shows that using pretrained transformer-based models as surrogate models substantially improve sample efficiency and hit recovery under strict labeling constraints. Motivated by this, we target the prediction of reduction potentials for TMCs using the UMA model.~\cite{UMA} While this model was trained on 100M DFT energies and forces across diverse charge and spin states, including TMCs, for use as a machine learning force field, studies with other MLFFs have shown that latent representations from such models are highly informative and transferable across different chemical prediction tasks.~\cite{kim2025leveragingneuralnetworkinteratomic} Leveraging this approach allows us to maximize data efficiency without sacrificing expressivity.

Namely, for each molecule in the candidate pool specified by atomic numbers $\{Z_i\}$, Cartesian coordinates $\{\mathbf{R}_i\}$, total charge $q$, and spin multiplicity $s$, we define the atom-wise representation as
\begin{equation}
\mathbf{h}_i^{l=0}(\{Z_j\}, \{\mathbf{R}_j\}, q, s)
\end{equation}
where $i$ indexes atoms and $l=0$ denotes the rotation-invariant (scalar) component of the final UMA layer before the prediction head. We then introduce a small adapter feed-forward neural network $f_{\theta}(\cdot)$, applied independently to each atomic representation. The transformed features are then passed to a final multilayer perceptron (MLP) $MLP_{\phi}(\cdot)$ and aggregated to produce the prediction $\hat{y}$:
\begin{equation}
    \hat{y} = \frac{1}{N} \sum_{i=1}^{N}MLP_{\phi}\!\left(
    f_{\theta}\!\left(
    \mathbf{h}_i^{l=0}(\{Z_j\}, \{\mathbf{R}_j\}, q, s)
    \right)
    \right)
\end{equation}
where $N$ denotes the total number of atoms in the molecule.
The model architecture is deliberately designed to leverage self-supervised pretraining for adapting molecular representations to the specific dataset, a practice shown to be beneficial in active learning settings when the same candidate pool is used for pretraining.~\cite{Masood2025} In our case, we train $f_{\theta}(\cdot)$ on the whole candidate pool utilizing the recently published LeJEPA~\cite{lejepa} directly on the embeddings obtained from UMA (see Methods). Notice that utilizing this pretraining strategy on already learned embeddings will not have the same representation learning benefits as applying it on the network producing said embeddings by noising the original input. Nevertheless this process is aimed at adapting the representation to the current domain without representation collapse, 
while preserving the information content of the underlying UMA atomic embeddings. 
During training on the labeled set this adapter component will not be frozen but a regularization will be applied so its weights do not diverge from the pretrained configuration, and it will otherwise be trained together with the MLP with an additional Dropout layer placed between the two. 

Prediction of the target properties is done without optimization of the geometries and considering the embedding of each molecule as was obtained during enumeration considering charge and spin multiplicity at its ground state. While geometry optimization may improve predictive performance, this pipeline is aimed at showing results under scalability and computational constraints. In this setting where the labels are assigned following a computational procedure, we only consider epistemic uncertainty. Given that the model is a lightweight, we can adopt ensemble uncertainty by training a different model per each of 10 different cross validation splits. Deep ensembles have been shown to yield well-calibrated uncertainty estimates and strong out-of-distribution detection performance, while remaining straightforward to implement and scalable.~\cite{lakshminarayanan2017simple,ovadia2019can}

\subsection{Efficiency of Chemical Space Exploration}\label{Sampling}

\begin{figure*}[ht!]
  \centering
  \includegraphics[width=\textwidth]{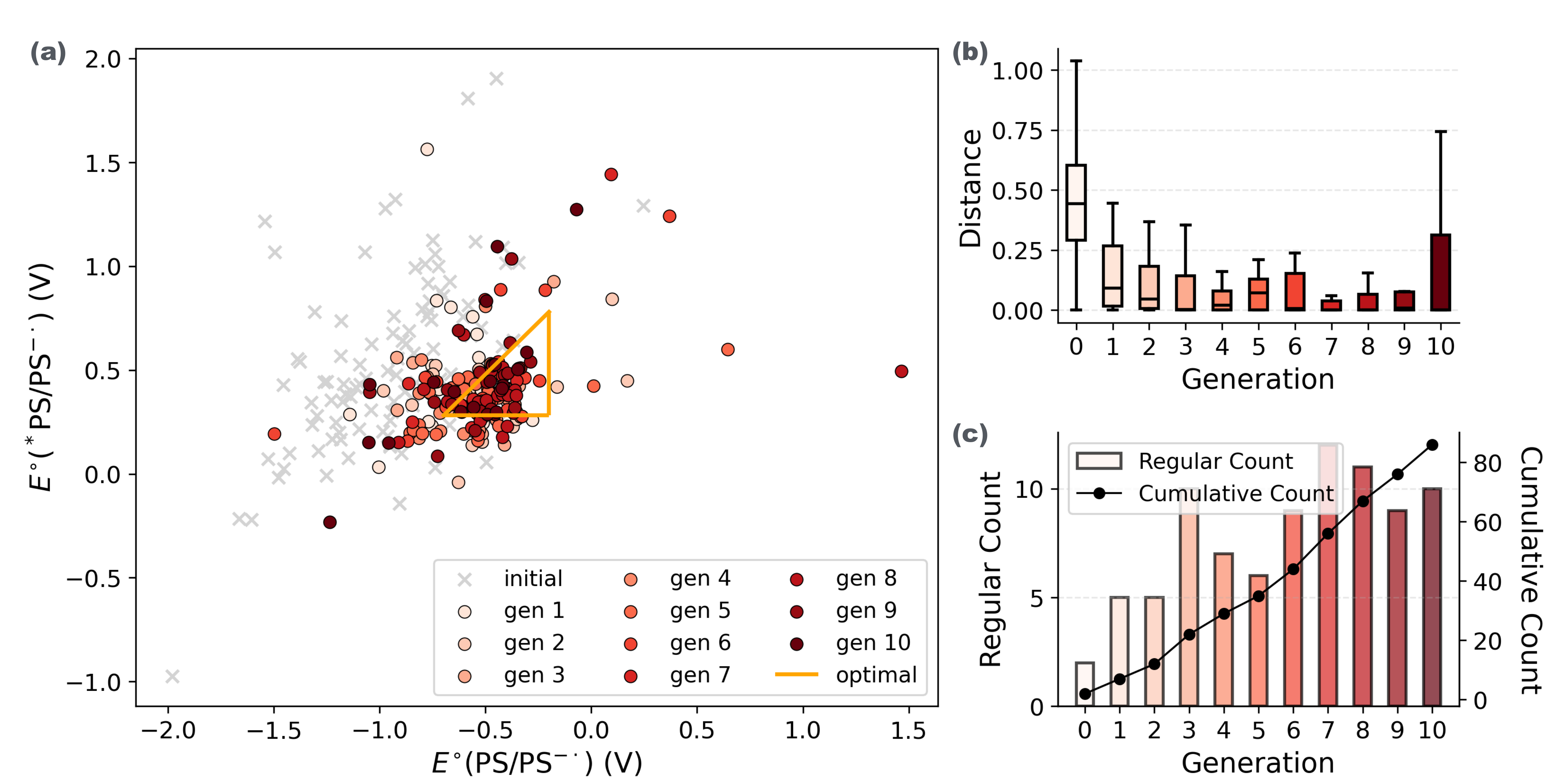}
  \caption{(a) Scatter plot in the $\big(E^\circ(\mathrm{PS}/\mathrm{PS}^{-\cdot}), E^\circ({}^*\mathrm{PS}/\mathrm{PS}^{-\cdot})\big)$ plane of the molecules selected by generation (\textit{gen}). The targeted region is shown as an orange boundary, whereas molecules in the initial training set are indicated by grey $\times$ markers. (b) Boxplot of the distance to the optimal region by generation. Generation 0 corresponds to the initial training set. Whiskers span the 5th–95th percentiles and the boxes represent the 25th–75th percentiles. (c) Bar plot showing the number of molecules sampled from the optimal region per generation, overlaid with the cumulative trend.}
  \label{fig:results}
\end{figure*}

The results obtained from the initial set of 100 TMCs are first examined in the $\big(E^\circ(\mathrm{PS}/\mathrm{PS}^{-\cdot})$, $ E^\circ({}^*\mathrm{PS}/\mathrm{PS}^{-\cdot})\big)$-plane. Analysis of this initial dataset shows that two molecules already fall within the target triangular region. Before analyzing the results obtained from the active learning (AL) pipeline, a set of 300 molecules was generated by uniform random sampling. This provides a qualitative estimate of the effectiveness of the AL search strategy relative to random exploration, as well as an indication of the difficulty of the search problem. Among these 300 randomly sampled molecules, only 8 fall within the optimal region, corresponding to an estimated probability of $2.67\%$ for sampling from the target region. The AL pipeline was then executed as described in the previous section for 10 generations, yielding a total of 300 evaluated TMCs, among which 200 were explored according to the model selection. The results are summarized in Fig.~\ref{fig:results}.

Figure~\ref{fig:results}a shows that the model quickly focuses its iterative selection process around a point inside the optimal region lying on its left side.  Overall, the sampled TMCs by generation get much closer to that region w.r.t. the original training set, as can be seen by the distribution of the distance values by generation in Fig.~\ref{fig:results}b, with most of the reduction happening in the first three or four generations. This trend is confirmed by the analysis of the number of molecules sampled inside the optimal region in Fig.~\ref{fig:results}c which shows that the number of optimal molecules sampled increases up to 12 per generation ($60\%$ sampling efficiency) remaining steadily above 9 per generation ($45\%$ sampling efficiency) after generation 6. Considering the cumulative number of molecules sampled within the optimal region, and including the initial dataset as part of the exploration process, 86 out of the 300 evaluated molecules fall inside the triangular region, corresponding to an overall sampling efficiency of $28.7\%$. Although this is only a coarse estimate, results point to a roughly 10-fold improvement in sampling efficiency with respect to random sampling.

\subsection{Chemical Design Principles for Type I PDT from Viable Complexes}\label{Viable}

\begin{figure*}[ht!]
  \centering
  \includegraphics[width=6.5in]{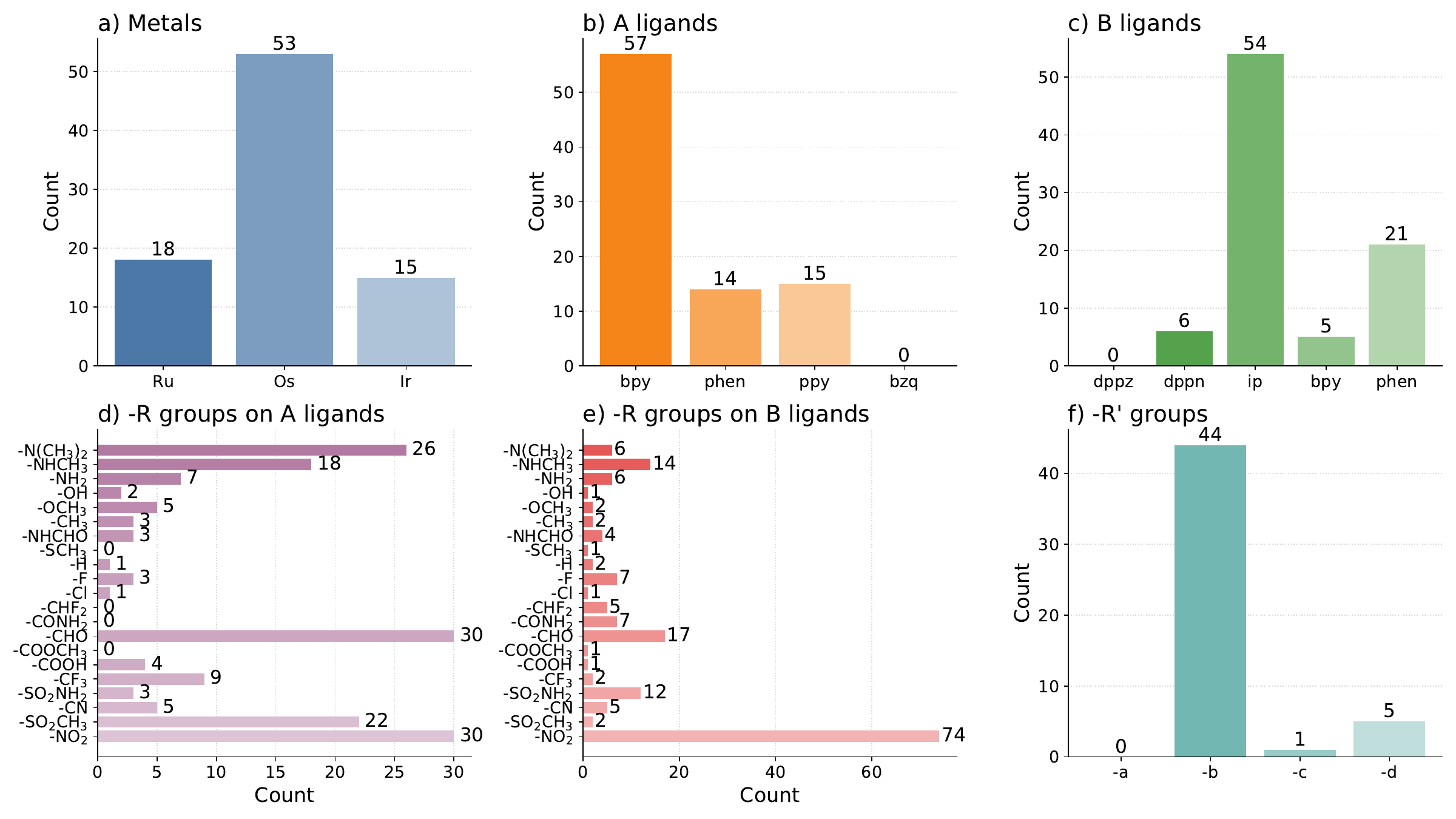}
  \caption{Distribution of structural components among the 86 complexes satisfying the design criteria. (a) Metal centers, (b) A ligands, (c) B ligands, (d) substituents (--R) on the A ligands, (e) substituents (--R) on the B ligands, and (f) substituents (--R') attached to the \textit{ip} B ligand. Bars represent the number of complexes containing a given structural element.}
  \label{fig:design_space_statistics}
\end{figure*}

Analysis of the identified viable complexes reveals clear chemical design principles governing Type~I photoreactivity in transition-metal photosensitizers. The 86 complexes satisfying the design constraints show consistent trends in the structural features favored by the active learning search. As shown in Fig.~\ref{fig:design_space_statistics}a, the distribution of metal centers shows a strong preference for Os(II) complexes (53 out of 86), followed by Ru(II) complexes (18) and Ir(III) complexes (15). This enrichment indicates that Os-based complexes more readily satisfy the combined constraints imposed by the objective function. Although spin--orbit coupling was not explicitly included as a design criterion, the preference for Os(II)-based complexes is nevertheless beneficial, since heavier metals are expected to promote efficient intersystem crossing and population of the triplet state, which is advantageous for reactive oxygen species generation in photodynamic therapy.

Among the optimal complexes, \textit{bpy} is the dominant A ligand (57 complexes), followed by \textit{ppy}, which appears specifically in the Ir subset (15 complexes). By comparison, \textit{phen} is less frequently selected (14 complexes), while \textit{bzq}, the corresponding Ir analogue, is not selected in the optimal set (0 complexes) (see Fig.~\ref{fig:design_space_statistics}b). Analysis of the --R substituents on the A ligands shows that the model preferentially selects substituents capable of substantially perturbing the ligand electronic structure, as given in Fig.~\ref{fig:design_space_statistics}d. On the donating side, strongly electron-donating groups (EDGs) such as --$\mathrm{N(CH_3)_2}$ (26 occurrences), --$\mathrm{NHCH_3}$ (18), and --$\mathrm{NH_2}$ (7) are frequently selected. Among electron-withdrawing substituents (EWGs), the distribution is dominated by strongly EWGs, particularly --$\mathrm{NO_2}$ (30), --$\mathrm{SO_2CH_3}$ (22), and --$\mathrm{CN}$ (5). Moderately EWGs are also common, including --$\mathrm{CHO}$ (30) and --$\mathrm{CF_3}$ (9). In contrast, substituents with a moderate EDG character, such as --$\mathrm{OH}$ (2) and --$\mathrm{OCH_3}$ (5), appear less frequently, while weakly perturbing groups near the electronic midpoint of the Hammett scale~\cite{hansch1991survey}—such as --$\mathrm{CH_3}$ (3) and --H (1)—are rare. Overall, the substituent distribution indicates that the model favors functional groups capable of producing substantial electronic perturbations of the ligand framework, thereby enabling precise positioning of the reduction potentials within the narrow thermodynamic window required for efficient Type~I electron transfer.

Analysis of the B ligand distribution (Fig.~\ref{fig:design_space_statistics}c) reveals a strong preference for the \textit{ip} scaffold, which dominates the set of viable complexes (54 occurrences). The \textit{phen} ligand is also significantly represented (21 occurrences), while \textit{dppn} (6) and \textit{bpy} (5) appear less frequently, and \textit{dppz} is absent. Substituents on the B ligand are strongly biased toward EWGs (see Fig.~\ref{fig:design_space_statistics}e). The distribution is dominated by the strongly withdrawing --$\mathrm{NO_2}$ group (74 occurrences), along with --$\mathrm{SO_2NH_2}$ (12) and --$\mathrm{CN}$ (5). Moderately EWGs, including --$\mathrm{CHO}$ (17), --$\mathrm{CONH_2}$ (7), and --$\mathrm{CHF_2}$ (5), are additionally represented, whereas the weakly EWG --$\mathrm{F}$ appears 7 times. Strong EDGs occur less frequently, but are still retained in several viable candidates, most notably --$\mathrm{NHCH_3}$ (14), --$\mathrm{N(CH_3)_2}$ (6), and --$\mathrm{NH_2}$ (6). Among the --R' groups attached to the \textit{ip} ligand, the \textit{b} motif is overwhelmingly preferred (44 occurrences), whereas \textit{d} appears 5 times, \textit{c} appears once, and \textit{a} is absent, as shown in Fig.~\ref{fig:design_space_statistics}f. Because these --R' substituents extend the conjugated system of the B ligand, their prevalence indicates that $\pi$-extension of the \textit{ip} scaffold, particularly through motif \textit{b}, assists in fine-tuning the excited-state redox properties required for efficient Type~I electron-transfer pathways.

Taken together, these observations define a set of practical design principles for Type~I PDT photosensitizers. Optimal candidates favor heavy metal centers, particularly Os(II). At the ligand level, strong electronic asymmetry is required, achieved through combinations of EDGs on the A ligand and EWGs on the B ligand, enabling precise tuning of excited state and redox properties. In addition, the use of strongly perturbing substituents is essential to access the narrow redox window, while $\pi$-extension of --R' group on the \textit{ip} B ligand scaffold, facilitates fine control of excited-state energetics. These features collectively define a structural motif for optimal candidates and provide a rational framework for the design of next-generation Type~I PDT photosensitizers.

\begin{figure*}[ht!]
  \centering
  \includegraphics[width=3.2in]{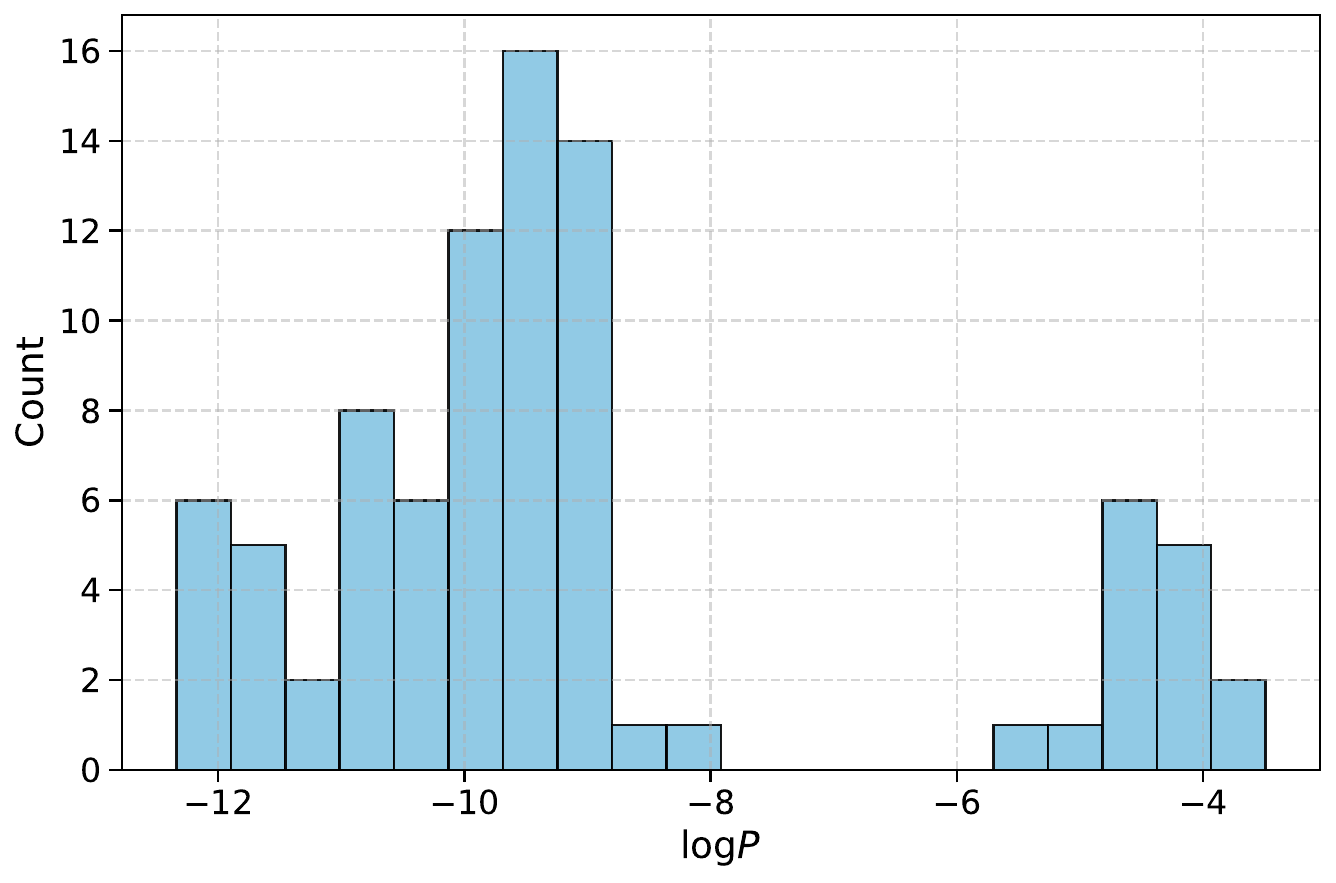}
  \caption{Distribution of calculated octanol–water partition coefficients (log$P$) for the 86 complexes that satisfy the design criteria.}
  \label{fig:logP_distribution}
\end{figure*}

In addition to satisfying the aforementioned design criteria, the octanol–water partition coefficients (log$P$) were calculated for the 86 viable complexes as a measure of lipophilicity. As shown in Fig.~\ref{fig:logP_distribution}, the resulting log$P$ values are shifted toward negative values, indicating pronounced hydrophilicity. The distribution clearly separates into two populations. The dominant population (log$P$ $\approx -12.5$ to $-8.0$) corresponds to Ru(II) and Os(II) complexes. Within this group, the lowest log$P$ values are associated with complexes containing a high density of strongly EWGs together with the \textit{ip} B ligand. At the upper end of this main population (log$P$ $\approx -8.9$ to $-8.3$), the complexes typically contain fewer EWGs and instead incorporate moderate or strong EDGs on the A ligand, while the B ligand is predominantly \textit{dppn}. A second population with significantly higher log$P$ values ($-5.5$ to $-3.5$) corresponds exclusively to Ir(III) complexes consistent with their lower molecular charge. At the lower end of this range, the complexes contain the \textit{ip} B ligand and the ligand framework remains dominated by strong EWGs. Toward the upper end of this population, the complexes instead contain the \textit{dppn} B ligand and incorporate mixtures of EWGs and EDGs across the ligand framework. The presence of such substituents, together with the more extended aromatic surface of the \textit{dppn} ligand, reduces overall polarity and leads to higher log$P$ values. Overall, these observations indicate that the log$P$ distribution is governed by the interplay of substituent polarity, B-ligand identity, and metal center.

\subsection{Model Analysis and Performance}

To assess the model's performance, several aspects were analyzed. First, we start by comparing predictions from two configurations: (i) a model whose adapter component was pretrained on the full candidate pool, and (ii) a model whose adapter was not pretrained. This comparison serves as a sanity check performed only at the first iteration, i.e., after training on the initial training set, to assess the impact of the pretraining step. This is done only on this initial step as it is the only case during iterations where the training set is not skewed towards the targeted optimal region. The analysis was conducted on the set of 300 independent random samples used to estimate the effectiveness of random sampling in the previous section.

\begin{figure}
    \centering
    \includegraphics[width=\linewidth]{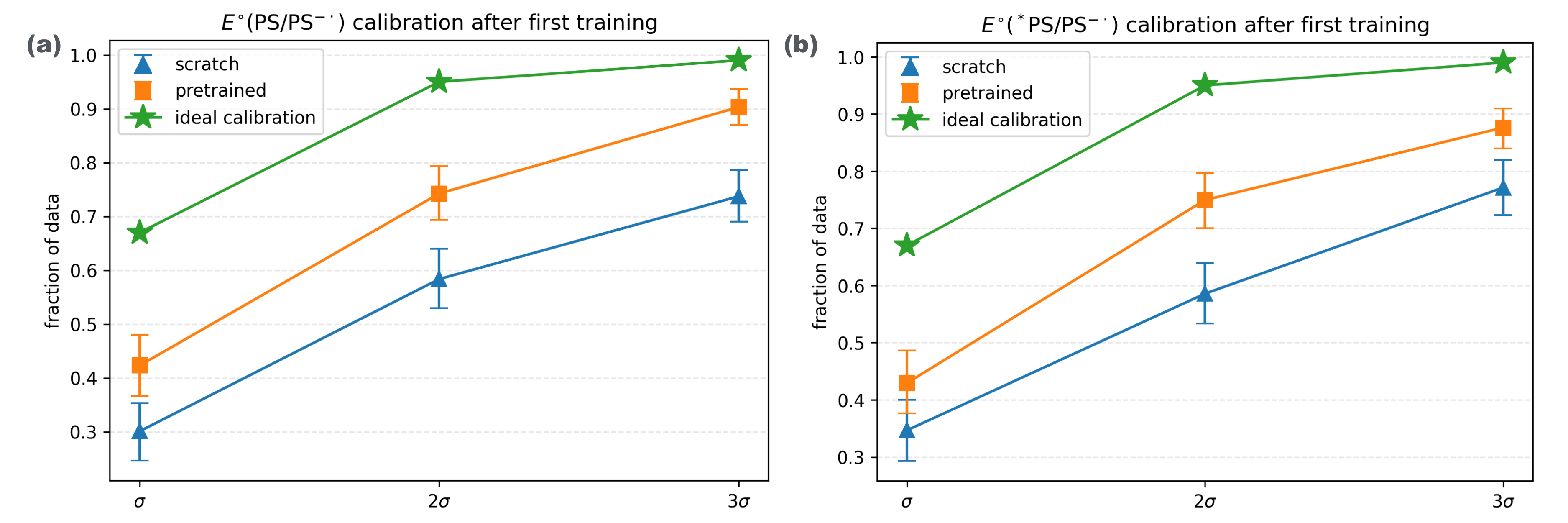}
    \caption{(a) Calibration plot for $E^\circ(\mathrm{PS}/\mathrm{PS}^{-\cdot})$ and (b) $E^\circ({}^*\mathrm{PS}/\mathrm{PS}^{-\cdot})$. Fractions of absolute deviations bellow one, two, and three standard deviations of the model ensemble are shown for both pretrained and non-pretrained (scratch) models. Ideal calibration values are included for comparison.}
    \label{fig:calibration}
\end{figure}

The results show that the overall Mean Absolute Error (MAE) is essentially the same in both cases for both properties. The non-pretrained model achieves a MAE of $0.187~\mathrm{V}$ with a 95\% confidence interval (CI) of $[0.169~\mathrm{V}, 0.204~\mathrm{V}]$ for $E^\circ(\mathrm{PS}/\mathrm{PS}^{-\cdot})$, and a MAE of $0.203~\mathrm{V}$ with a 95\% CI of $[0.183~\mathrm{V},0.222~\mathrm{V}]$ for $E^\circ({}^*\mathrm{PS}/\mathrm{PS}^{-\cdot})$. The pretrained model achieves a MAE of $0.205~\mathrm{V}$ with a 95\% CI of $[0.185~\mathrm{V},0.225~\mathrm{V}]$ for $E^\circ(\mathrm{PS}/\mathrm{PS}^{-\cdot})$, and an MAE of $0.202~\mathrm{V}$ with a 95\% CI of $[0.183~\mathrm{V},0.221~\mathrm{V}]$ for $E^\circ({}^*\mathrm{PS}/\mathrm{PS}^{-\cdot})$. 
A more pronounced and decisive difference emerges when comparing uncertainty calibration. Although both pretrained and non-pretrained models exhibit overconfidence on this set, as shown in Fig.~\ref{fig:calibration} the pretrained model is closer to the ideal calibration w.r.t. its non-pretrained counterpart which is more overconfident. These results reflect that while the information in the frozen UMA embeddings mainly drive performance (no noticeable difference in MAE), the pretraining of the adapter component on the whole candidate pool allows for a better initial model calibration. Mean values and confidence intervals for both MAE and calibration were estimated via bootstrapping with 1000 resamples.

\begin{figure}
    \centering
    \includegraphics[width=\linewidth]{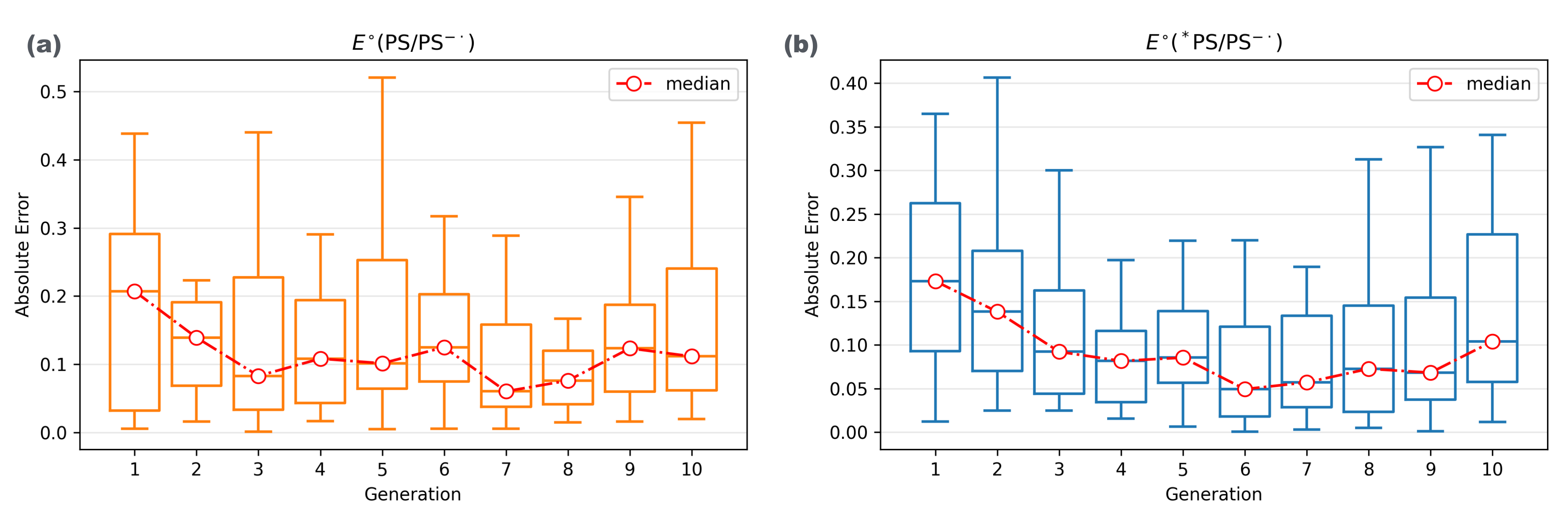}
    \caption{(a) Boxplot of the absolute errors distribution for $E^\circ(\mathrm{PS}/\mathrm{PS}^{-\cdot})$ on the 20 selected molecules for each of the 10 training iterations (namely generation $N$ considers the predictions on molecules from generation $N$ obtained with the models trained on data from generations up to $N-1$). The trend of the median value is also reported. (b) Corresponding results for $E^\circ({}^*\mathrm{PS}/\mathrm{PS}^{-\cdot})$.}
    \label{fig:AE_trend}
\end{figure}

Regarding the model performance as a function of training iteration, we considered the predictions obtained on the 20 selected molecules at each iteration. The performance in terms of boxplots of the Absolute Errors (AEs) is shown in Fig.~\ref{fig:AE_trend}. The results are noisy, which is expected given the small batch sizes and an acquisition function that prioritizes uncertainty when breaking ties. Nevertheless, the median shows a mildly descending trend for both $E^\circ(\mathrm{PS}/\mathrm{PS}^{-\cdot})$ and $E^\circ({}^*\mathrm{PS}/\mathrm{PS}^{-\cdot})$, with most of the error reduction happening within the first 3 to 5 generations. As a general consideration, one should not expect drastic improvements in model accuracy here as we are still in a regime of exceedingly low data with respect to the candidate pool, while using an exploitation-first acquisition function. 

Further considerations can be made by referring to the results reported in the~\ref{Sampling} and~\ref{Viable} subsections. Figure~\ref{fig:results}b shows that the distribution of distances from the optimal region improves consistently across generations, becoming increasingly skewed toward zero, with the final iteration maintaining a low median despite an overall increase in distribution spread. Also, considering the analysis provided in Fig.~\ref{fig:design_space_statistics}, despite the fact that the initial set contained only two Ru(II) complexes in the optimal region, the model was able to generalize to a diverse set of optimal compounds with different chemistries.

\section{Conclusion and Perspective}

In this work, we establish a data-efficient active learning–driven framework for the targeted discovery of transition-metal photosensitizers for Type~I PDT. By combining a chemically structured design space with selective quantum-chemical evaluations and pretrained atomic representations (UMA),~\cite{UMA} we show that efficient navigation of a multi-million compound space can be achieved with only 300 calculations. Relative to random sampling, the active learning strategy dramatically improves recovery of Type~I--suitable complexes, converting an otherwise intractable high-throughput screening problem into a scalable, mechanism-aware discovery framework. 

Analysis of the identified complexes reveals clear chemical design principles governing Type~I photoreactivity. Efficient candidates occupy a constrained redox window and preferentially combine heavy metal centers, particularly Os(II), with ligand environments that introduce strong electronic asymmetry through combinations of electron-donating and electron-withdrawing substituents. These features enable precise tuning of ground- and excited-state reduction potentials while maintaining favorable physicochemical properties, providing actionable guidelines for experimental design. Additionally, a defining strength of this framework is the direct prediction of ground- and excited-state reduction potentials—key descriptors governing electron-transfer reactivity. Because excited-state redox potentials cannot be measured directly and are typically inferred from electrochemical and spectroscopic data,~\cite{Tucker2012} their reliable computational prediction provides a powerful tool for experimental prioritization. In addition to PDT, these quantities are foundational to photocatalysis,~\cite{prier2013visible} photoredox organic synthesis,~\cite{reischauer2021emerging} CO$_2$ reduction,~\cite{ulmer2019fundamentals} and water splitting,~\cite{nishioka2023photocatalytic} positioning this strategy at the interface of photomedicine and sustainable energy research.

The current predictive ceiling is set by the computational protocol used to generate the training labels, which depends on both the chosen electronic structure method and the treatment of the chemical environment. To enable large-scale screening, the present workflow adopts several practical approximations. Solvent effects are treated with an implicit solvation model that neglects specific solute--solvent interactions that may influence photophysical behavior; explicit solvent models could therefore improve predictive fidelity. Likewise, although effective core potentials account for core electrons, the present protocol does not employ fully relativistic Hamiltonians that become increasingly important for heavier transition metals.~\cite{demissie2015dft} In addition, approximate exchange--correlation functionals such as B3LYP, although computationally efficient, can fail in systems with significant multireference character and near-degenerate electronic states.~\cite{nandy2021computational} Higher-level multireference approaches offer improved accuracy but remain computationally prohibitive at the scale required for large chemical space exploration.~\cite{Szalay2011,lyakh2012multireference} Quantum computing provides a potential long-term route to address these challenges by enabling more efficient treatment of strongly correlated electronic systems,~\cite{cao2019quantum,bauer2020quantum} although practical applications remain in development and may become tractable in the foreseeable future.~\cite{zehr2025quantum}
From the ML standpoint, frozen UMA embeddings proved to be the main driver of performance. While this pragmatic choice maximizes efficiency and scalability across the large chemical space, we also observe modest improvements in predictive accuracy after the initial training rounds. Future iterations could explore more refined techniques and alternative pretraining strategies to further improve on this aspect.

Crucially, the pipeline's architecture presented here is fully method-agnostic: any electronic structure approach capable of delivering redox energetics—whether higher-level multireference methods or future quantum-computing-based solvers—can be seamlessly integrated into the data acquisition loop. Moreover, while the present study employs a mechanistically defined redox feasibility objective, the framework is inherently flexible. Alternative objective functions can be readily implemented to enable expanded multiobjective optimization tailored to specific biomedical or catalytic contexts. In particular, future iterations of the discovery pipeline could benefit from objective functions that incorporate other measurable determinants of PDT performance. While the present framework focuses on ground and excited state redox energetics to define mechanistic feasibility for electron transfer photochemistry, the effectiveness of a photosensitizer in biological systems ultimately depends on a broader set of photophysical, photochemical, and biological properties. Experimentally accessible parameters such as absorption within the therapeutic window, molar extinction coefficients, excited state lifetimes, photostability under irradiation, and the efficiency of reactive oxygen species generation all influence photodynamic activity.~\cite{munegowda2022ru} In addition, biological factors including cellular uptake, subcellular localization, and phototoxicity under normoxic and hypoxic conditions play critical roles in determining therapeutic efficacy. Incorporating experimentally measured observables such as singlet oxygen quantum yields, ground and excited state redox potentials, oxygen consumption rates, and phototoxic dose response relationships could enable the construction of hybrid objective functions that more closely reflect real photodynamic performance.

Such experimentally guided optimization would also enable iterative refinement of the model through closed loop discovery in which computational predictions inform targeted synthesis and photobiological evaluation and the resulting experimental data are incorporated back into the learning pipeline. By integrating experimental feedback with data efficient active learning strategies, future implementations of this framework could accelerate the discovery of transition metal photosensitizers with improved photochemical reactivity and biological performance, particularly for challenging environments such as hypoxic tissue and conditions in which oxygen depletion during photodynamic treatment further limits therapeutic efficacy.

In conclusion, we strongly believe that the work presented herein constitutes a significant advancement in the rational, data-efficient discovery of transition metal complexes with targeted functionality. Concerning the specific application, our approach will help shed new light on the development of next-generation TMC photosensitizers for PDT,~\cite{ma2025current} photocatalysis,~\cite{prier2013visible} and solar energy conversion.~\cite{tran2012recent} More broadly, by uniting mechanism-driven objective design, scalable active learning, and future-ready electronic structure integration, this framework establishes a transformative paradigm for navigating complex transition metal chemical space with unprecedented efficiency and predictive power.

\section{Methods}
\subsection{DFT Calculation Details}

All quantum-chemical calculations were performed with ORCA 6.1.1.~\cite{neese2020orca} The computational workflow was automated using an in-house–modified version of the ORCA Python Interface (OPI).~\cite{FACCTsOPIv102025} The molecular geometries were fully optimized in the gas phase using the GFN2-xTB method.~\cite{bannwarth2019gfn2} The frequency analysis calculations were performed at the same level of theory to confirm the nature of all stationary points as minima (no imaginary frequencies) and to obtain the zero-point vibrational energies (ZPVE) and the Gibbs free energies in the gas phase ($\Delta G_\text{gas}^\circ$). Single-point electronic energies were subsequently computed on the xTB-optimized geometries using the hybrid B3LYP exchange-correlation functional.~\cite{Becke1988,Lee1988,beck1993density} The def2-SVP basis set was employed for all atoms,~\cite{weigend2005balanced} together with the corresponding def2/J auxiliary basis set for density fitting.~\cite{weigend2006accurate} For Ru, Os, and Ir, the def2 effective core potentials (def2-ECPs) were employed.~\cite{leininger1996accuracy} Open-shell doublet and triplet states were treated within the unrestricted Kohn–Sham formalism. Solvent effects were incorporated using the conductor-like polarizable continuum model (CPCM)~\cite{barone1998quantum} with default parameters as implemented in ORCA with water as the solvent. 

The ground-state reduction potentials, $E^\circ(\text{PS/PS}^{-\cdot})$, were obtained from the solution-phase Gibbs free energy ($\Delta G_\text{soln}^\circ$) differences between the reduced and neutral PS species ($\mathrm{PS} + e^- \rightarrow \mathrm{PS}^{-\cdot}$). The solution-phase Gibbs free energy for individual species were obtained by combining the gas-phase free energies ($\Delta G_\text{gas}^\circ$) derived from the xTB along with the electronic energies ($E_\text{elec}$) and the solvent free energies ($\Delta G_\text{solv}^\circ$) as obtained with B3LYP single-point calculations with the CPCM model ($\Delta G_\text{soln}^\circ = \Delta G_\text{gas}^\circ + E_{\text{elec}} + \Delta G_\text{solv}^\circ$).  The resulting solution-phase free energy change was converted to a redox potential according to $E^\circ(\text{PS/PS}^{-\cdot})=-\Delta G^\circ(\text{PS/PS}^{-\cdot})/nF$ where $n$ is number of transferred electrons and $F$ is the Faraday constant. All potentials were referenced to the SHE in water using an absolute potential of 4.281~V.\cite{kelly2006aqueous,kelly2007single} Excited-state reduction potentials $E^\circ(^*\text{PS/PS}^{-\cdot})$ were calculated using the Rehm–Weller relationship~\cite{rehm1970kinetics} in which the excited-state potential is derived from the ground-state reduction potential and the adiabatic $\text{T}_1 \rightarrow \text{S}_0$ transition ($E^\circ(^*\text{PS/PS}^{-\cdot}) = E^\circ(\text{PS/PS}^{-\cdot}) + \Delta E_{0-0}(\text{T}_1/\text{S}_0)$). The $\Delta E_{0-0}(\text{T}_1/\text{S}_0)$ energy was determined from the B3LYP energies in the water solvents on the xTB-optimized $\text{S}_0$ and $\text{T}_1$ geometries and by including ZPVE calculated at the xTB level.~\cite{adamo2013calculations} Finally, the log$P$ values were derived from the difference in Gibbs free energies of solvation in water and n-octanol, computed with the CPCM model at the B3LYP level on the xTB-optimized geometries. 

\subsection{Construction of Database Details}

Ligand generation and molecular assembly were implemented in a series of Jupyter notebooks with Python version 3.14.2 using RDKit~\cite{RDKit2025} (version 2025.09.3). Enumeration was performed using a custom Python Enumerator class developed specifically for this workflow. The \textit{bpy} scaffold was defined as a Simplified Molecular Input Line Entry System (SMILES)~\cite{weininger1988smiles,weininger1989smiles} and used to query PubChem.~\cite{kim2025pubchem} The retrieved structures were filtered with a predefined SMILES Arbitrary Target Specification (SMARTS) pattern, that specifies neutral nitrogen donor atoms without additional substituents to have zero charge and adjacent carbon atoms belonging to exactly one ring. From this filtered set of $N$ remaining molecules we carry out a scaffold decomposition aimed at removing all --R groups while maintaining the unique core structures containing our \textit{bpy} pattern. From the remaining cores, \textit{bpy} and \textit{phen} were selected to construct the A ligands and \textit{dppz}, \textit{dppn} and \textit{ip} were chosen to build the B ligands. Except for predefined attachment sites (Fig.~\ref{fig:DB_strategy}), substitution positions were restricted to hydrogen atoms. We used the selected cores to decompose the molecules matching those patterns into core --R group fragments, yielding a list of substituents associated with each selected core scaffold which we refer to as --R' for the fragment attachment to the \textit{ip} B ligand and --R for the rest. The lists of --R and --R’ substituents were refined by additional filtering to include the desired fragments shown in Fig.~\ref{fig:DB_strategy}, with manual additions of functional groups via atom-mapped SMILES. The final sets of ligands are assembled combinatorially resulting in a set of $N$ A ligands and $M$ B ligands. Final A and B ligands were stored separately as CSV files and serialized as canonical isomeric SMILES to ensure deterministic representation. Complex assembly was performed by first generating 3D geometries for each of the considered ligands once using RDKit, then by positioning them according to the coordination geometry of TLD-1433, using the nitrogen atoms bonded to the metal center and the two carbon atoms connecting them as anchor points. Alignment of the ligands to such anchor points was performed using Kabsch algorithm. For building the complex with \textit{ppy} and \textit{bzq} ligands, one of the coordinating nitrogen atoms was replaced with a negatively charged carbon center. The assembly was repeated with Os(II) and Ir(III) as the transition metal center, resulting in an ASE database (SQLite backend) containing $2~170~434$ complexes.

\subsection{Training and Model Details}

The adapter network $f_{\theta}(\cdot )$ mentioned in main text is chosen to have two 256 dimensional hidden layers with layer normalization placed before the activation function, and a final layer reducing the dimensionality back to the original 128 dimensions from the UMA embeddings. After a dropout layer with $p = 0.1$, the adopted $MLP_{\phi}(\cdot )$ has one 64 dimensional hidden layer before prediction. All nonlinearities here considered are Sigmoid-Weighted Linear Unit (SiLU) functions.

Pretraining was applied exclusively to $f_{\theta}(\cdot)$. For each molecule, 10 stochastic views were generated from the UMA-embedding-based atomic representations. Each view was constructed by randomly masking atomic features with probability $0.25$. The masked atomic representations were then processed through $f_{\theta}(\cdot)$, and the resulting atomic outputs were aggregated via mean pooling.
Formally, the representation corresponding to view $v$ is given by
\begin{equation}
\mathbf{g}_{\theta}^{v}(\mathbf{x})
=
\frac{1}{N}
\sum_{i=1}^{N}
f_{\theta}\!\left(
\mathbf{x}_v
\right)
\end{equation}
where $N$ denotes the number of atoms in the molecule, $v \in \{1,\dots,10\}$ indexes the independently noised views, and where we consider $\mathbf{x}$ to be the UMA embeddings of size $N\times 128$. This pretraining optimizes the loss function from \cite{lejepa}, which we will rewrite here for our setting:
\begin{equation}
\mathcal{L}_{\text{pretrain}} = \lambda\sum_{v=1}^{10}\rm{SIGReg}(\{\mathbf{g}_{\theta}^{v}(\mathbf{x})_n\}_{n \in B}) + \frac{1}{B}\sum_{n = 1}^{B}\frac{1}{10}\sum_{v = 1}^{10}||\mathbf{g}_{\theta}^{v}(\mathbf{x})_n - \mathbf{\mu}_{n}||^2
\end{equation}
where $\lambda = 0.01$, $n$ runs over the training batch $B$, $\mathbf{\mu_n = \frac{1}{V}\sum_{v = 1}^{10}{g}_{\theta}^{v}(\mathbf{x})_n}$, and SIGReg is the Sketched Gaussian Isotropic Regularization as described in Ref.~\citenum{lejepa}.
This regularization term is based on a differentiable statistical test which pushes the distribution of the learned mean embeddings to be close to an isotropic Gaussian in embedding space. While in the original work this has also the function of building global representations that minimize empirical risk, since we do predictions from atomic representations in the downstream task here we only exploit it to avoid feature collapse.
The pretraining utilized the whole candidate pool where 80\% was used for training and the remaining 20\% was used for validation in order to have a metric to save the best model by. This procedure was run on $4$ NVIDIA A100 GPUs, with a global batch size of $800$ and utilizing a learning rate triangular scheduler oscillating between $10^{-5}$ and $10^{-3}$ with a half-period of $2000$ steps. The optimizer used is AdamW with a weight decay parameter of $10^{-4}$. The pretraining was run for $24 \rm{h}$ for a total of $200$ epochs.

At each active learning iteration, training is carried out using the adapter component obtained during pretraining. A loss term with weight $0.1$ is added to constrain divergence from the pretrained weights while allowing flexibility to adapt to downstream task. This is otherwise trained together with the final $MLP_{\phi}(\cdot)$, preceded by a dropout layer in input as mentioned initially. The training loss is a standard Mean Squared Error (MSE) loss on the two predicted properties. The ensemble was trained using 10-fold cross-validation, where each model was fitted on a different leave-one-out partition of the data. We utilized a batch size of 25, and a learning rate triangular scheduler oscillating between $5^{-6}$ and $5^{-4}$ with a half-period of $20$ steps (note that for the initial 100 datapoints we have 4 gradient steps per epoch). This is run until an early stopping call is triggered with a patience of 50 epochs saving the best model according to minimum validation loss. All trainings are run on an NVIDIA A100 GPU. As training times differ by iteration, it can only be reported that training all 10 models in each iteration takes roughly 10 minutes, whereas the inference step (evaluating 10 model predictions on the remaining $\sim 2M$ candidates and computing the acquisition function) takes roughly 20 minutes.

\subsection{Active Learning Pipeline Details}
In this section we provide additional details on the AL pipeline. Starting from the enumerated database --- an ASE database with SQLite backend handled via \texttt{fairchem}~\cite{Shuaibietal2024}'s \texttt{ASEDBDataset} class --- we generate a dataset of atomistic representations using UMA~\cite{UMA}. This is implemented as a customized \texttt{PyTorch Geometric}~\cite{Feyetal2025, FeyLenssen2019} \texttt{OnDiskDataset} with a \texttt{RocksDB} backend, where each molecule is stored as a \texttt{PyTorch Geometric} datapoint containing an $N_{\mathrm{atoms}}\times 128$ feature tensor. Since the molecular graph information is already encoded in the UMA node embeddings, no explicit graph connectivity is stored. At runtime, training and inference are controlled by storing only the indices of molecules assigned to the training set, validation set, or candidate pool, allowing the foundation model to be run only once for the full database. All subsequent training and inference within the AL loop relies exclusively on our lightweight model ensemble, ensuring scalability and fast inference times.

The clustering based selection of the initial candidate set, as well as the clustering for the diversification of the iterative selection is carried out with the Batched K-Means algorithm as implemented in \texttt{scikit-learn} with a batch size of 5000. 

\begin{acknowledgement}
Work on “Quantum Computing for Photon-Drug Interactions in Cancer Prevention and Treatment” is supported by Wellcome Leap as part of the Q4Bio Program. We acknowledge EuroHPC for granting access to the Leonardo supercomputing facility at CINECA (Italy) under projects EHPC-REG-2025R05-171 and EUHPC-R05-221. L.N. thanks the Research Council of Finland for financial support through the Finnish Quantum Flagship, project number 358878.
\end{acknowledgement}
\newline

\noindent\textbf{Author contributions}: A.F. and F.P. conceived the project and developed methodology. A.F., P.H., and F.P. implemented the software workflow. A.F., J.E., and M.S. implemented dataset construction protocol and generated the training dataset. A.F., P.H, L.N., and F.P. performed benchmark calculations and acquired the data. A.F., M.S., and F.P. prepared the figures. A.F., J.E., S.M., M.S, and F.P. wrote the original draft. All authors were involved in data analysis, and reviewing and editing the manuscript. F.P. supervised the project.
\\


\noindent\textbf{Conflict of interest}\\
The authors declare no conflict of interest.

\linespread{1}\selectfont
\bibliography{references}

\end{document}